\documentclass[superscriptaddress,nofootinbib]{revtex4}

\usepackage{graphicx}
\usepackage{dcolumn}
\usepackage{bm}
\usepackage{hyperref}
\usepackage{textcomp}
\usepackage{amsmath}
\usepackage{multirow}
\usepackage{color}

\begin{document}




\title{Search for modulations of the solar {$^7$Be} flux\\ in the next-generation neutrino observatory LENA}

\author{Michael Wurm}\email[Corresponding author, e-mail:~]{mwurm@ph.tum.de}
\affiliation{Physik-Department, Technische Universit\"at M\"unchen, 85748 Garching, Germany}
\author{Barbara Caccianiga}
\author{Davide D'Angelo}
\affiliation{Dipartimento di Fisica, Universit\`a degli Studi e INFN, 20133 Milano, Italy}
\author{Stefano Davini}
\affiliation{Dipartimento di Fisica, Universit\`a e INFN, Genova 16146, Italy}
\author{Franz von Feilitzsch}
\author{Marianne G\"oger-Neff}
\affiliation{Physik-Department, Technische Universit\"at M\"unchen, 85748 Garching, Germany}
\author{Tobias Lachenmaier}
\affiliation{Kepler Center, Universit\"at T\"ubingen, 72076 T\"ubingen, Germany} 
\author{Timo Lewke}
\affiliation{Physik-Department, Technische Universit\"at M\"unchen, 85748 Garching, Germany}
\author{Paolo Lombardi}
\author{Livia Ludhova}
\affiliation{Dipartimento di Fisica, Universit\`a degli Studi e INFN, 20133 Milano, Italy}
\author{Quirin Meindl}
\affiliation{Physik-Department, Technische Universit\"at M\"unchen, 85748 Garching, Germany}
\author{Emanuela Meroni}
\author{Lino Miramonti}
\affiliation{Dipartimento di Fisica, Universit\`a degli Studi e INFN, 20133 Milano, Italy}
\author{Randolph M\"ollenberg}
\author{Lothar Oberauer}
\author{Walter Potzel}
\affiliation{Physik-Department, Technische Universit\"at M\"unchen, 85748 Garching, Germany}
\author{Gioacchino Ranucci}
\affiliation{Dipartimento di Fisica, Universit\`a degli Studi e INFN, 20133 Milano, Italy}
\author{Marc Tippmann}
\author{J\"urgen Winter}
\affiliation{Physik-Department, Technische Universit\"at M\"unchen, 85748 Garching, Germany}

\date{\today}

\begin{abstract}
A next-generation liquid-scintillator detector will be able to perform high-statistics measurements of the solar neutrino flux. In LENA, solar {$^7$Be} neutrinos are expected to cause 1.7$\times$10$^4$ electron recoil events per day in a fiducial volume of 35 kilotons. Based on this signal, a search for periodic modulations on sub-percent level can be conducted, surpassing the sensitivity of current detectors by at least a factor of 20. The range of accessible periods reaches from several minutes, corresponding to modulations induced by helioseismic g-modes, to tens of years, allowing to study long-term changes in solar fusion rates. 
\end{abstract}

\maketitle



\section{Introduction}

\noindent Since the early days of solar neutrino flux measurements, efforts have been made to discover a variation of the neutrino rate over time. The most prominent effect is the experimentally confirmed annual variation of the flux by 7\,\% that is induced by the eccentricity of the Earth's orbit (e.\,g.\,\cite{ran07a,sk08sol}). But a variety of alternative sources of flux modulation is conceivable: The survival probability of solar electron neutrinos might be influenced by fluctuations of the solar matter density \cite{mir01,akh02,cha05} or by traversing terrestrial matter \cite{car86,bah97,deg01} before reaching the detector on Earth. Solar neutrino production rates might even change in the course of the solar cycle of about 11 years \cite{kra90}, or might be subject to short-term variations correlated to the oscillation of the solar core temperature induced by helioseismic waves \cite{sno09}.

The next-generation neutrino observatory LENA will be a low-background liquid-scintillator experiment \cite{mar06,mar08,wur10det}. In recent years, the potential of this detector technique has been demonstrated by the first real-time measurement of {$^7$Be} neutrinos in the solar neutrino experiment Borexino \cite{bx07be7}. Opposed to the comparatively low rate of $\sim$50 counts per day (cpd) in Borexino, about $10^4$\,cpd of {$^7$Be} neutrino events are expected in LENA due to the considerably larger target mass (Sect.\,\ref{SecNeuRat}). Given this large statistics, it is clear that LENA will be much more sensitive to temporal flux fluctuations as any of the preceding solar neutrino experiments, including the Super-Kamiokande detector \cite{sk05sm}. In the following analysis based on the Lomb-Scargle method \cite{lom76,sca82} (Sect.\,\ref{SecLomSca}), we demonstrate that LENA will be sensitive to fluctuations on a sub-percent level over a wide range of frequencies: Studies of the sensitivity cover modulation periods from several minutes up to tens of years (Sect.\,\ref{SecAnaRes}). The resulting discovery potential for a number of plausible modulation sources will be presented in Sect.\,\ref{SecDiscus}.

\section{{$^7$Be} Neutrino Rate and Backgrounds}
\label{SecNeuRat}

The low background environment necessary for a low-energy rare-event search is reflected in the design of the future LENA detector: The target volume will encompass about 5$\times$10$^4$\,m$^3$ of organic liquid scintillator contained in an inner cylindrical nylon vessel of 100\,m height and 26\,m diameter. Scintillation light is observed by photomultiplier tubes (PMTs) that are mounted to a scaffolding in 2\,m distance from the nylon cylinder: About 45,000 8''-PMTs equipped with light concentrators would be needed to reach the intended photocoverage of 30\,\%. Additional 2$\times$10$^4$\,m$^3$ of non-scintillating buffer liquid will surround the scintillator, providing  shielding against the $\gamma$ rays emitted by the PMTs and the surrounding steel tank of 30\,m diameter. Outside the tank, an external water volume will serve as an active muon veto at the detector perimeter, while the lid of the tank will be covered either by plastic scintillator panels or resistive plate chambers (RPCs) to enhance the spatial resolution of muon tracking. Further details are given in \cite{wur10det}.

Depending on the natural radioactivity content of the detector materials used, the fiducial volume for solar neutrino detection will have to be chosen considerably smaller than the full scintillator volume. Monte Carlo simulations show that the $\gamma$ ray background expected in LENA will be dominated by the natural radioactivity of the PMT glass in case of a steel tank. The total rate of $\gamma$ rays emitted by {$^{40}$K} and elements of the {$^{238}$U} and {$^{232}$Th} decay chains is $\sim$170\,kBq. However, the self-shielding effect of the organic liquid reduces the rate of $\gamma$ rays reaching the bulk of the detector considerably.

Tab.\,\ref{TabShiFid} shows the fiducial mass $M_\mathrm{fid}$ depending on the radius $r_\mathrm{fid}$ of the chosen cylindric volume. By scaling the {$^7$Be}-$\nu$ rate of 49 counts per day and 100\,t observed in Borexino, one can derive the total expected rate ($3^\mathrm{rd}$ column in Tab.\,\ref{TabShiFid}) as function of $M_\mathrm{fid}$ that linearly scales with the available volume. However, {$^7$Be}-$\nu$ detection in liquid scintillator relies on the detection of electron recoils \cite{bx07be7}. Part of these will lie below the detection threshold of 250\,keV, corresponding to $\sim$43\,\% of the total recoil spectrum\footnote{The exact values depend on the energy resolution of the detector, that defines the smearing of the endpoint of the {$^{14}$C} spectrum.}. Therefore, the signal rate remaining inside the observation window from 250-800\,keV must be compared to the rate of $\gamma$ rays reaching the fiducial volume and creating a signal of the same visible energy. In the following, a fiducial mass of $\sim$35\,kt is chosen, corresponding to $\sim$10$^4$ {$^7$Be}-$\nu$ recoil events and $\sim$300 $\gamma$ events per day inside the detection window ($4^\mathrm{th}$ and $5^\mathrm{th}$ column).

\begin{table}[h]
\begin{center}
\begin{tabular}{|cc|c|cc|}
\hline
\multicolumn{2}{|c|}{Fiducial Volume} & Rate [cpd] & \multicolumn{2}{c|}{{$^7$Be} window [cpd]} \\
$r_\mathrm{fid}$\,[m] & $M_\mathrm{fid}$\,[kt]& {$^7$Be}-$\nu$ & {$^7$Be}-$\nu$ & BG($\gamma$) \\
\hline
13 & 43.8 & 21.5$\times$$10^4$ & 12.3$\times$$10^4$ & 1.2$\times$$10^6$ \\
12 & 36.6 & 17.9$\times$$10^4$ & 10.2$\times$$10^4$ & 6.8$\times$$10^3$ \\
11.5 & 33.2 & 16.3$\times$$10^4$ & 9.34$\times$$10^3$ & 2.5$\times$$10^2$ \\
11 & 30.1 & 14.7$\times$$10^4$ & 8.42$\times$$10^3$ & 7.8$\times$$10^0$ \\
10.5 & 27.1 & 13.2$\times$$10^4$ & 7.56$\times$$10^3$ & 5$\times$$10^{-1}$ \\
10 & 24.3 & 11.9$\times$$10^4$ & 6.82$\times$$10^3$ & 3$\times$$10^{-2}$ \\
\hline
\end{tabular}
\caption{The radius $r_\mathrm{fid}$ of the fiducial volume in LENA as well as the mass contained ($M_\mathrm{fid}$) depends on the required shielding against the $\gamma$ background produced by the PMTs. Signal and background rates are computed for LAB (C$_{18}$H$_{30}$, $\rho=0.86$\,g/$\ell$) \cite{lab-spec}. The middle column lists the total rates of {$^7$Be}-$\nu$ events as function of $M_\mathrm{fid}$, while the last columns present the rates of {$^7$Be}-$\nu$ events and external $\gamma$-rays expected in the {$^7$Be} detection window from 250 to 800\,keV.}
\label{TabShiFid}
\end{center}
\end{table}

The experience with Borexino shows that the radiopurity requirements of $10^{-17}$\,g of {$^{238}$U} and {$^{232}$Th} per gram of liquid scintillator (or $\sim$0.5 decays per day and ton) are achievable, which is a prerequisite for the detection of the {$^7$Be} signal \cite{bx08be7}. As the necessary precautions in the production, transport and handling of the organic solvents as well as potent purification procedures have been put to test in both the Counting Test Facility (CTF) and Borexino itself, similar or even lower levels of radioactivity can be expected in LENA. In all of the following analysis, it is assumed that the radiopurity conditions achieved in Borexino will be reproduced for LENA.

\begin{figure}
\centering
\includegraphics[width=0.485\textwidth]{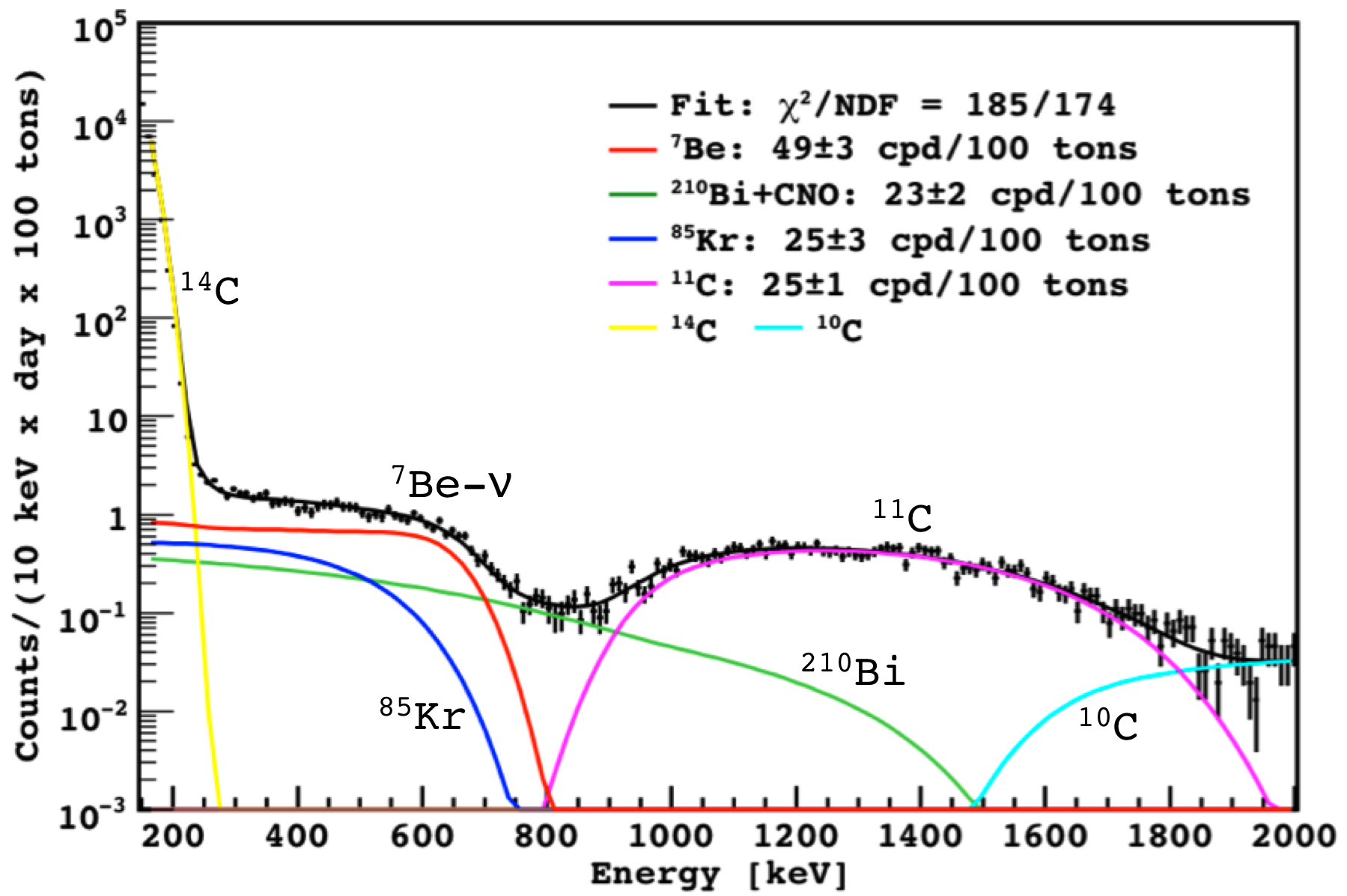}
\caption{Solar neutrino recoil spectra from elastic electron scattering as measured by Borexino \cite{bx08be7}. The data points correspond to the measured spectrum below 2\,MeV for 15\,kton$\cdot$days of exposure. Both signal and background spectra are included in the fit. {$^{210}$Po} has been subtracted using $\alpha/\beta$ discrimination. 
}
\label{FigSolSpe}
\end{figure}

Recent developments in the R\&D of organic solvents indicate that the optical transparency of the liquid to the scintillation light will surpass former expectations \cite{wur10sca}. Some brands of linear alkylbenzene (LAB) feature attenuation lengths on the scale of 20\,m \cite{SNO08phd}, corresponding to a light collection efficiency of about 450\,pe/MeV (photoelectrons per MeV) in LENA. This value is only slightly worse than the value achieved in Borexino \cite{bx08be7}. This is a significant improvement, as self-absorption of the scintillation light in the liquid is a major issue and yields of $\sim$100\,pe/MeV would be expected if the technology of Borexino was simply duplicated. Due to the enhanced light collection, the low-energy electron-recoil spectrum of LENA might look much the same as that observed in Borexino (Fig.\,\ref{FigSolSpe}). Therefore, analysis techniques based on a fit of pre-calculated signal and background spectra to the overall sum spectrum as they are currently used in Borexino are expected to perform equally well in LENA \cite{bx07be7,bx08be7}. 

\section{Lomb-Scargle Periodigrams}
\label{SecLomSca}

\begin{figure}
\centering
\includegraphics[width=0.485\textwidth]{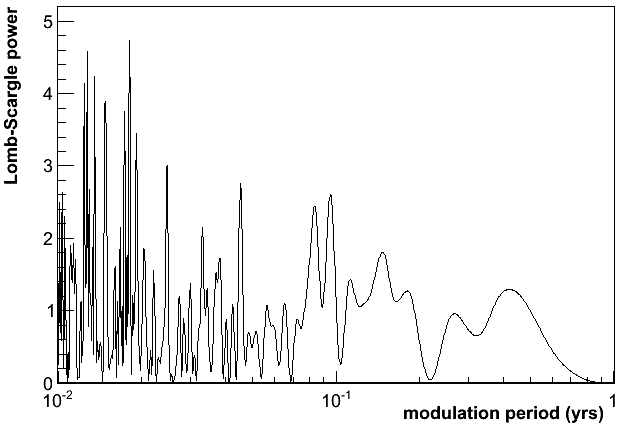}~
\includegraphics[width=0.485\textwidth]{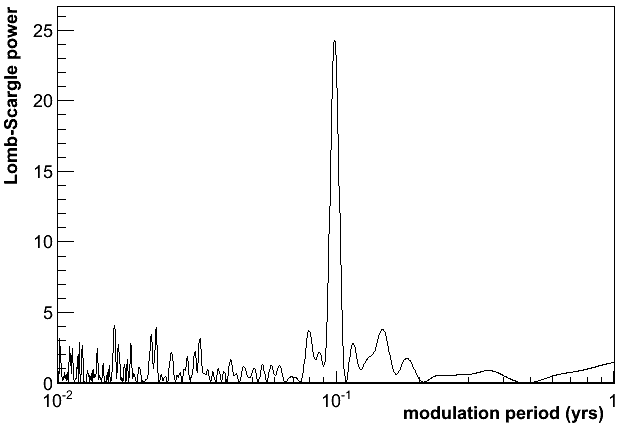}
\caption{Lomb-Scargle periodigrams for a MC data set of 2 years measurement time. \textit{Left:} White noise spectrum. \textit{Right:} A modulation of 2\,\% relative amplitude and a period of 0.1 years was included. A corresponding peak is visible at the expected position, that is also clearly exceeding the white noise level.}
\label{FigLomPer}
\end{figure}

A variety of different techniques has been applied for modulation search in solar neutrino data, reaching from unbinned maximum likelihood methods \cite{sno05, ran07b} to wavelet approaches sensitive to transient modulations \cite{ran07a}. In the following Lomb-Scargle (LS) periodigrams are used \cite{lom76, sca82, sno05, gno05, ran07a}, a common method to analyze a binned data set for periodical modulations of the type
\begin{equation} \label{EqBasMod}
N(t) = N_0\cdot\left(1+A\cdot\sin(2\pi t/T+\varphi)\right).
\end{equation}
Here, $N(t)$ is the expected event rate at time $t$, while $N_0$ represents the mean rate, $A$ indicates the relative amplitude of the modulation, $T$ describes its period and $\varphi$ the phase relative to the start of the measurement. The power $P$ at which a modulation can be found in the data is obtained by weighting the difference between the number of events $N(t_i)$ in every data bin $i$ and the expected mean value $N_0$ with cosine and sine functions that oscillate with the investigated period $T$:
\begin{eqnarray}
P = \frac{1}{2\sigma^2}\left( \frac{\left[\sum_{i=1}^n w_i(N(t_i)-N_0)\cos\left(2\pi \frac{t_i - t_p}{T}\right)\right]^2 }{\sum_{i=1}^n w_i\cos^2\left(2\pi \frac{t_i - t_p}{T}\right)} + \frac{\left[\sum_{i=1}^n w_i(N(t_i)-N_0)\sin\left(2\pi \frac{t_i - t_p}{T}\right)\right]^2 }{\sum_{i=1}^n w_i\sin^2\left(2\pi \frac{t_i - t_p}{T}\right)}\right)
\end{eqnarray}
where $n$ is the number of bins, $\sigma$ the standard deviation of $N$, and $t_i$ the time at which the data corresponding to bin $i$ was acquired \cite{sno05}. As the quadratic sums of both cosine and sine are used, the result is independent of the modulation phase as long as the modulation period is short in comparison to the overall measurement time. The weights $w_i$ allow to compensate differences in the bin width $w$ of individual data samples. As the analysis presented in Sect.\,\ref{SecAnaRes} will be based on equally wide bins (neglecting dead time), $w_i=w$ has been used. Therefore, all weight factors cancel each other. The phase factor $t_p$ can be derived from the equation \cite{sno05}:
\begin{eqnarray}
\tan\left(4\pi \frac{t_p}{T}\right) = \frac{\sum_{i=1}^N w_i\sin\left(4\pi \frac{t_i}{T}\right)}{\sum_{i=1}^N w_i\cos\left(4\pi \frac{t_i}{T}\right)}.
\end{eqnarray}
Fig.\,\ref{FigLomPer} shows on the left a LS periodigram of a white noise Monte Carlo (MC) data set, while on the right a modulation of $A=2$\,\% and $T=0.1$\,yrs is included. The LS power $P$ of a given modulation is primarily a function of its amplitude $A$. Statistical fluctuations of the bin content, $\Delta=\Delta N/N$, will alter both the maximum $P$ generated by white noise and the exact value observed for the actual modulation. To assess the significance of a modulation discovery, it is therefore necessary to know the statistical fluctuations of both the white noise level and the signal height.

\begin{figure}
\centering
\includegraphics[width=0.485\textwidth]{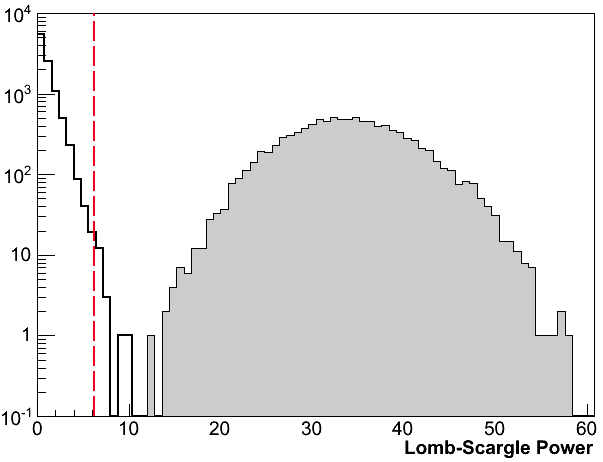}
\caption{Distribution of Lomb-Scargle powers of white-noise (solid line) and of a modulation of 2\,\% relative amplitude and 0.1 years period (shaded area) (compare Fig.\,\ref{FigLomPer}). There is a clear separation between the distributions. The sensitivity threshold $s$ is set at 99.7\,\% of the white noise distribution (dashed line).}
\label{FigPerSig}
\end{figure}

The analyses conducted in the next chapters are based on a common basic scheme that was adopted from \cite{sno05}: To determine whether a modulation of relative amplitude $A$ and period $T$ is pronounced enough to be detected, 10$^4$ MC data sets containing a modulation parameterized according to Eq.\,(\ref{EqBasMod}) are simulated, and the LS power $P$ for the signal at period $T$ is computed. Then, a white-noise sample of $10^4$ Monte Carlo (MC) data sets is generated for comparison. The detection threshold is set above 99.7\,\% of the white-noise LS power distribution, which corresponds to a 3$\sigma$ discovery potential. If all of the signal spectrum is above this detection threshold, the probability for a 3$\sigma$ discovery of the modulation corresponds to 100\,\% (compare Fig.\,\ref{FigPerSig}). In case there is an overlap and the signal distribution extends below the detection threshold, the discovery probability is reduced to the proportion of the distribution remaining above threshold. In the following analyses, the sensitivity threshold $s$ in LENA is defined as the amplitude for which the probability for a $3\sigma$ discovery is 90\,\%. 

\section{Results on Sensitivity}
\label{SecAnaRes}

The results presented here are based on a LS study of the expected signal rate generated by {$^7$Be}-$\nu$ electron recoils in LENA. The investigation will be presented in three regions defined by the modulation period $T$: Starting from values of $T$ of the order of days to years in which LENA features the largest sensitivity, very short periods of about 30\,min and very slow modulations on the scale of centuries are explored. A central issue in all analyses is the presence of radioactive background that superimposes the neutrino signal. The strategy to isolate the signal will differ depending on the chosen bin width $w$ and will be discussed in the individual sections.

\subsection{Medium-Term Modulations}
\label{SecMedMod}

First, the dependence of the detection sensitivity on measurement time and bin-content uncertainties will be outlined. Then the uncertainty level that can be expected from the {$^7$Be}-$\nu$ analysis will be discussed, relying on the experience gathered in Borexino. Finally, the resolution of neighboring modulation lines in the LS periodigram will be discussed.

\subsubsection*{Primary Parameters Governing Sensitivity}

The Lomb-Scargle method is most sensitive for modulations in an intermediate range of periods: $T$ should not significantly exceed the overall measurement time $\tau$, but should also not be so short to come into conflict with the bin width $w$. MC tests show that the sensitivity is independent of $T$ and also of the modulation phase $\varphi$ if these conditions are fulfilled (see also Fig.\,\ref{FigOptPer}). This aspect will be discussed in Sect.\,\ref{SecShoMod}.

One of the main factors determining the minimum detectable amplitude $A$ is the measurement time $\tau$. Moreover, the relative uncertainty of the event number in each bin, $\Delta =\Delta N/N$, has a large influence. To demonstrate these dependences, Tab.\,\ref{TabYrsUnc} shows LENA's sensitivity as a function of $\tau$ and $\Delta$. The chosen bin width $w=1$\,day. As stated in Sec.\,\ref{SecLomSca}, $s$ corresponds to the minimum amplitude that is discovered at 3$\sigma$ level in 90\,\% of all MC data sets. Samples of 10\,000 MC data sets each were used to determine the detection threshold by comparison of white noise and signal spectra.

\begin{table}
\begin{center}
\begin{tabular}{|c|cccccccccc|}
\hline
 & \multicolumn{10}{c|}{Relative uncertainty per bin $\Delta=\Delta N/N$\,[\%]} \\
$\tau$ [yrs]& 1 & 2 & 3 & 4 & 5 & 6 & 7 & 8 & 9 & 10 \\
\hline
1 & 0.34 & 0.66 & 0.97 & 1.29 & 1.61 & 1.93 & 2.25 & 2.57 & 2.92 & 3.21 \\
2 & 0.26 & 0.47 & 0.68 & 0.91 & 1.14 & 1.36 & 1.59 & 1.80 & 2.03 & 2.26 \\
3 & 0.19 & 0.38 & 0.57 & 0.75 & 0.93 & 1.19 & 1.29 & 1.48 & 1.66 & 1.84 \\
4 & 0.19 & 0.34 & 0.49 & 0.65 & 0.80 & 0.97 & 1.12 & 1.28 & 1.45 & 1.60 \\
5 & 0.18 & 0.29 & 0.45 & 0.58 & 0.72 & 0.86 & 1.08 & 1.15 & 1.29 & 1.43 \\
6 & 0.17 & 0.28 & 0.40 & 0.53 & 0.66 & 0.79 & 0.92 & 1.05 & 1.18 & 1.31 \\
7 & 0.17 & 0.28 & 0.39 & 0.49 & 0.61 & 0.73 & 0.85 & 0.98 & 1.09 & 1.20 \\
8 & 0.15 & 0.25 & 0.36 & 0.49 & 0.57 & 0.69 & 0.80 & 0.91 & 1.02 & 1.14 \\
9 & 0.14 & 0.26 & 0.36 & 0.44 & 0.54 & 0.65 & 0.80 & 0.92 & 0.97 & 1.15 \\
10 & 0.11 & 0.21 & 0.31 & 0.41 & 0.50 & 0.62 & 0.71 & 0.81 & 0.91 & 1.02 \\
\hline
\end{tabular}
\caption{Sensitivity $s$ of modulation search as a function of measurement time $\tau$ and of the relative uncertainty of the rate measurement $\Delta = \Delta N/N$. $10^4$ MC data sets were generated. The sensitivity $s$ corresponds to the minimum amplitude (in \%) that would be observed at 3$\sigma$ level for 90\,\% of the MC data sets.}
\label{TabYrsUnc}
\end{center}
\end{table}

\subsubsection*{Uncertainties from the Fit}

To assess the expected sensitivity $s$ in LENA, it is necessary to determine both statistic and systematic uncertainties of the neutrino-rate measurement during a period corresponding to the bin width $w$. As the {$^7$Be}-$\nu$ rate surpasses all other neutrino rates by at least one order of magnitude, these neutrinos are the most sensitive probes for this analysis \footnote{The question whether {$^8$B} neutrinos might be another promising candidate for measurements of solar temperature fluctuations will be discussed in Sect.\,\ref{SubsHelMod}.}. Based on a fiducial mass of 35\,kt, the expected {$^7$Be} neutrino rate will be of the order of $10^4$\,cpd. Consequently, a statistical error of 1\,\% can be expected for a standard bin width $w=1$\,d.

However, the data collected in Borexino suggests that there are several sources of radioactive background inside the {$^7$Be} detection window. The most prominent ones are the decays of {$^{210}$Po}, {$^{210}$Bi}, and {$^{85}$Kr} dissolved in the scintillator.

The isotope {$^{210}$Po} emits $\alpha$-particles with an energy of 5.41\,MeV. Due to the quenching effects present in liquid scintillators, this line is shifted to about 450\,keV in the electron recoil-spectrum (depending on the specific scintillator composition) \cite{bx07be7, bx08be7}. Assuming for LENA a replication of the initial radioactive background levels of Borexino, the contamination with {$^{210}$Po} would be considerable, in the beginning surpassing the expected {$^7$Be} rate by more than two orders of magnitude. Despite this large rate, the effect of {$^{210}$Po} on the {$^7$Be} rate determination can be assumed as negligible: As demonstrated in Borexino, the efficiency of liquid-scintillator detectors to identifiy $\alpha$ events by a pulse-shape analysis reaches levels of at least 98\,\% \cite{bx07ab}. Moreover, the {$^{210}$Po} half-life of 138\,d is short compared to the expected operation time of LENA. The initial value of {$^{210}$Po} contamination will be reduced each year by more than a factor of 6 as long as it is not fed by the decay of long-lived {$^{210}$Pb}. 

The ultimate limit for the {$^{210}$Po} reduction by decay is set by the {$^{210}$Pb} contamination of the scintillator, which produces {$^{210}$Po} by the beta decay chain ${^{210}\mathrm{Pb}}\rightarrow{^{210}\mathrm{Bi}}\rightarrow{^{210}\mathrm{Po}}$. In Borexino, the initial amount of {$^{210}$Po} by far surpassed the concentrations of {$^{210}$Bi} and {$^{210}$Pb}; the {$^{222}$Rn} chain is out of equilibrium. However, {$^{210}$Pb} with a half-life of 22.3\,yrs is relatively long-lived. Even assuming an operation time of 30 years for LENA, the final {$^{210}$Pb} will still be at about 40\,\% of the start value. While the $\beta$ spectrum of the {$^{210}$Pb} decay itself features a sufficiently low endpoint of 64\,keV not to be in conflict with neutrino detection, its comparatively short-lived decay daughter {$^{210}$Bi} features a $\beta$ endpoint of 1162\,keV, severely interfering with the signals of {$^7$Be}, pep, and CNO neutrinos. As the emitted particle is an electron it cannot be distinguished from a neutrino-induced electron recoil by its pulse shape. The only possibility is an identification of this contribution to the overall recoil spectrum by its known spectral shape. Similar arguments can be made in the case of {$^{85}$Kr}, a $\beta$-unstable isotope with an endpoint energy of 687\,keV and a half-life of 10.8\,yrs.

For an estimate of the impact of these backgrounds and other systematic sources on the overall uncertainty in LENA, we refer to the results of Borexino. The {$^7$Be} neutrino analysis presented in \cite{bx08be7} is based on a data set corresponding to 198 live days of data taking or about 7\,500 neutrino events. This corresponds to 1/2 of the statistics collected in LENA in a single day. As described before, the {$^7$Be} neutrino rate for the overall period was determined by a spectral fit to the data, taking into account the recoil spectrum of {$^7$Be} and the contribution of {$^{85}$Kr} and {$^{210}$Bi} background events. For Borexino, the resulting {$^7$Be} event rate is $49\pm3_\mathrm{(stat)}\pm4_\mathrm{(syst)}$\,cpd. The uncertainties are mostly caused by the determination of the fiducial mass and by the detector response function.

Unlike the uncertainty of the absolute {$^7$Be} neutrino rate, a relative rate measurement as in the case of the modulation search will not be affected by the exact knowledge of the fiducial mass. On the other hand, the uncertainty in the detector response function enters the analysis by ambiguities in the spectral fit. It can be significantly reduced by calibrations based on the insertion of radioactive sources in the detector. Both Borexino and Kamland have undertaken dedicated campaigns. Based on the result of the calibrations, the Borexino collaboration is now aiming at a reduction of the uncertainties on the {$^7$Be} measurement, corresponding to an uncertainty of $\sim$3\,\% for the {$^7$Be} rate extracted from spectral fits and an overall uncertainty of $\sim$5\,\% including other systematics. In principle, the uncertainty obtained in LENA for a single-day measurement might be even lower than this value: The average background rate per day can be determined using the statistics corresponding to time spans of months or years, and the remaining uncertainty on the fiducial mass can be neglected. Excluding a change in background rates due to novel contaminations during the measurement time, the level of radioactive contamination of the scintillator is constant (ignoring for the moment the decay of radioimpurities). Using the mean background rates as pull parameters for the spectral fit will further reduce the uncertainty of the day-by-day determination of the {$^7$Be} rate. 

Conservatively, an uncertainty of $\Delta = \Delta N/N = 3\,\%$ can be expected for the individual bins, corresponding to 10$^4$ events as the bin width is a day. Assuming a measurement duration and a stability of the fluctuation of 1\,(10)\,yr(s), a sensitivity of about 1\,\%\,(0.3\,\%) can be reached for medium-term fluctuations (compare Tab.\,\ref{TabYrsUnc}). 

\subsubsection*{Resolution of Modulation Period}

It is possible that the solar neutrino flux is subject to multiple superimposed modulations, featuring either a whole spectrum of frequencies or several discrete values. In the latter case, the ability to distinguish between nearby lines in the LS periodigram and the determination of the statistical uncertainty in the peak frequency are of great interest:

The minimum difference in modulation period that is needed for resolving two close-by superimposed modulations is determined by the corresponding line width $\Delta T$ in the LS periodigram. In Fig.\,\ref{FigLomPer}, the relative width $\Delta T/T$ is 2\,\% (1$\sigma$) for a modulation of $T=0.1$\,yrs and $\tau=2$\,yrs. 

MC calculations show that both the measurement uncertainty per bin $\Delta$ and the modulation amplitude $A$ have no influence, at least as long as $A>s$ (Tab.\,\ref{TabYrsUnc}). Merely the measurement time $\tau$ and the period $T$ alter the resulting $\Delta T$. Varying both parameters, $\Delta T$ proves to be inversely proportional to $\tau$. At the same time, it scales as the square of the period $T^2$. This behavior is equivalent to the assumption that only the number of available measurement bins $n = T/w = T/(\tau/T) =T^2/\tau$  has an influence on the accuracy. Assuming a bin width of 1 day, $\Delta T/T$ can therefore be written as:
\begin{eqnarray}
\frac{\Delta T }{T}(T,\tau) = 0.02\left(\frac{T}{0.1\,\mathrm{yr}}\right)^2\left(\frac{2\,\mathrm{yr}}{\tau}\right) = 4\left(\frac{T}{1\,\mathrm{yr}}\right)^2 \left( \frac{1\,\mathrm{yr}}{\tau}\right).
\end{eqnarray}
In case of a single modulation, the mean value of the line, i.\,e.\,the modulation period $T$, can be determined at high accuracy from the LS periodigram (Fig.\,\ref{FigLomPer}). To determine the jitter in the maximum value of the line, periodigrams of 100 MC data sets were created according to the parameters mentioned above. The peak value featured a root mean square of 6$\times$10$^{-4}$ relative to $T$.

\subsection{Long-term Changes in Flux}
\label{SecLonMod}

The LS method is most sensitive to the discovery of flux oscillations for periods shorter than the measurement time. The currently planned operation time of the LENA detector is 10 to 30\,yrs, which is of the order of magnitude of a full solar cycle. However, it is not excluded that there might be modulations of the total nuclear energy production of the Sun on longer time spans which would be reflected in the emitted neutrino flux. Therefore, a series of MC calculations was performed to determine the sensitivity to very long-term modulations dependent on the period $T$.

\begin{figure}
\centering
\includegraphics[width=0.485\textwidth]{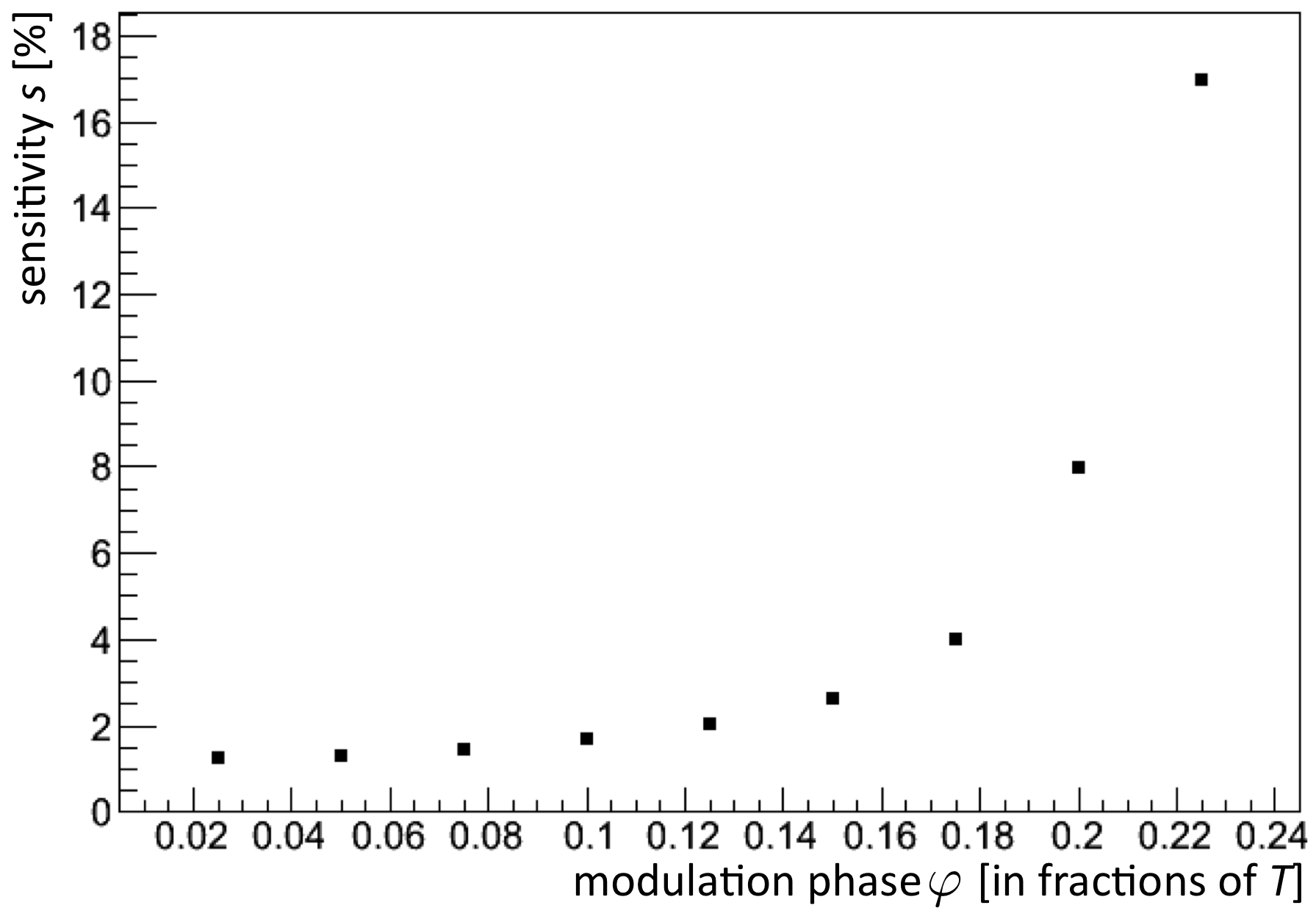}
\caption{Influence of the modulation phase $\varphi$ on the sensitivity $s$ to a modulation of $T=200$\,yrs, that varies by an order of magnitude ($\tau=10$\,yrs, $\Delta=3\,\%$).}
\label{Fig200Yrs}
\end{figure}

Different to the situation for medium-term modulations, the sensitivity to slow oscillations also is a function of the modulation phase $\varphi$ relative to the start of the measurement period. For $\tau\ll T$, the visible change in detected neutrino rate will be smallest near the flux maxima and minima, $N=N_0(1\pm A)$, while it will be largest near the zero points of the oscillation function, $N\approx N_0$. This expectation is confirmed by MC calculations displayed in Fig.\,\ref{Fig200Yrs}: For a measurement time $\tau=10\,$yrs and $\Delta= \Delta N/N =3\,\%$, the minimum detectable amplitude $A$ of a modulation with a period of 200\,yrs varies between 1.3\,\% and more than 17\,\%, depending on $\varphi$.

\begin{figure}
\centering
\includegraphics[width=0.485\textwidth]{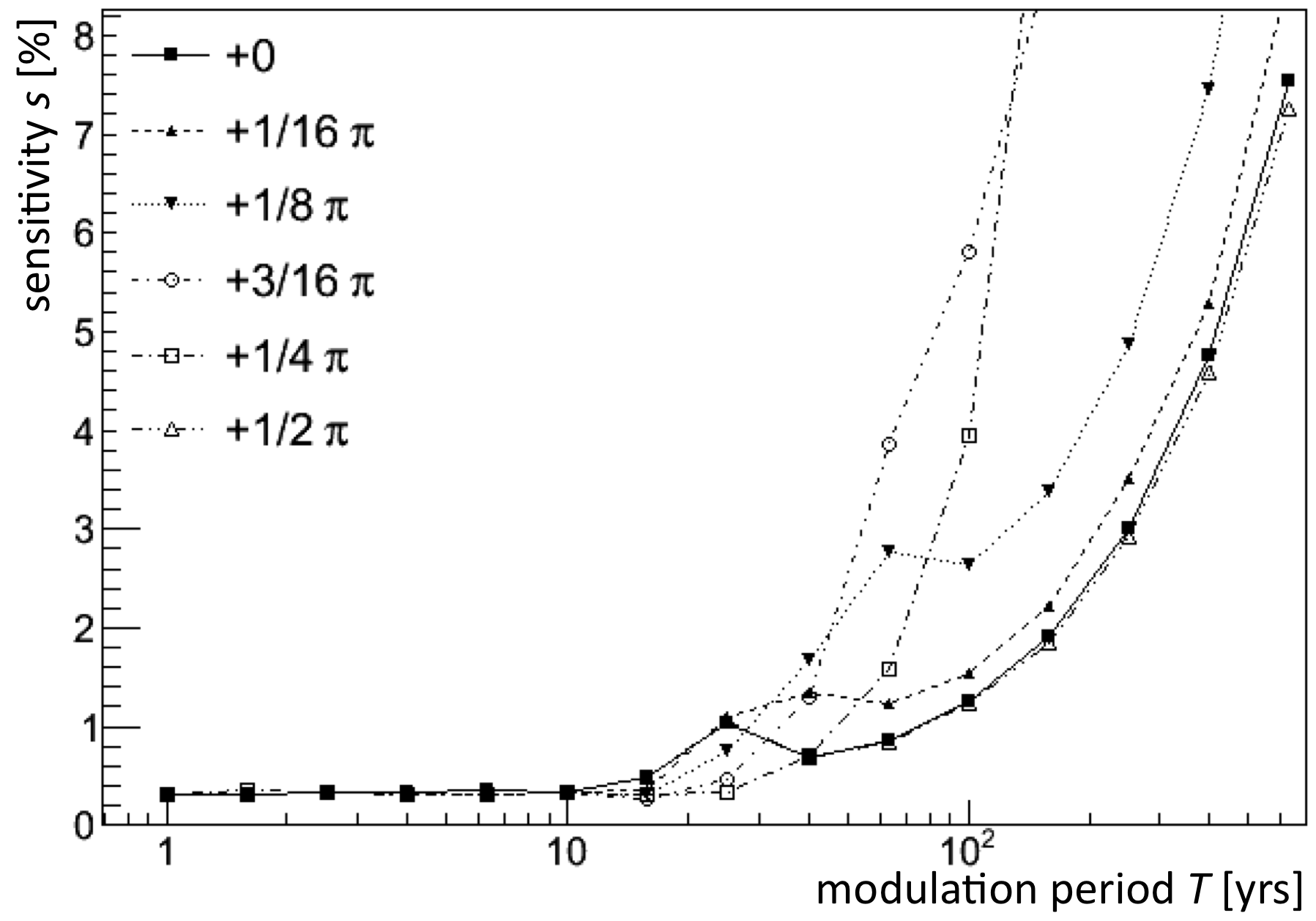}
\caption{Influence of the modulation phase $\varphi$ on the sensitivity $s$ for modulations with periods longer than the measurement time, $T>\tau$ ($\tau=10$\,yrs, $\Delta=3\,\%$).}
\label{FigLngYrs}
\end{figure}

The predicted sensitivity depends on the actual combination of $\tau$, $T$, and $\varphi$. This is illustrated in Fig.\,\ref{FigLngYrs} that for $\tau=10$\,yrs scans the range of modulation periods from 10 to 500\,yrs for various values of $\varphi$. For $T>2\tau$, the phase $\varphi$ becomes the dominant factor in determining the achievable sensitivity, while for $T>5\tau$ a disadvantageous value of $\varphi$ is able to obscure a modulation that would otherwise be well visible. The longer the operation time of the detector, the more these effects will be reduced.

\subsection{High-Frequency Modulations}
\label{SecShoMod}

While the main concern for LS modulation search at low frequencies is the total measurement time $\tau$, the central issue for very high frequencies is the bin width $w$ and the corresponding uncertainty on the bin contents $\Delta(w)$. As discussed in Sect.\,\ref{SecLomSca}, there are certain techniques as e.\,g.\,unbinned maximum-likelihood methods that do not suffer from the the finite bin width \cite{sno05}. But also in this case, statistical subtraction of the background will limit the achievable sensitivity. 

\subsubsection*{Minimum Length of Modulation Period }

\begin{figure}
\centering
\includegraphics[width=0.485\textwidth]{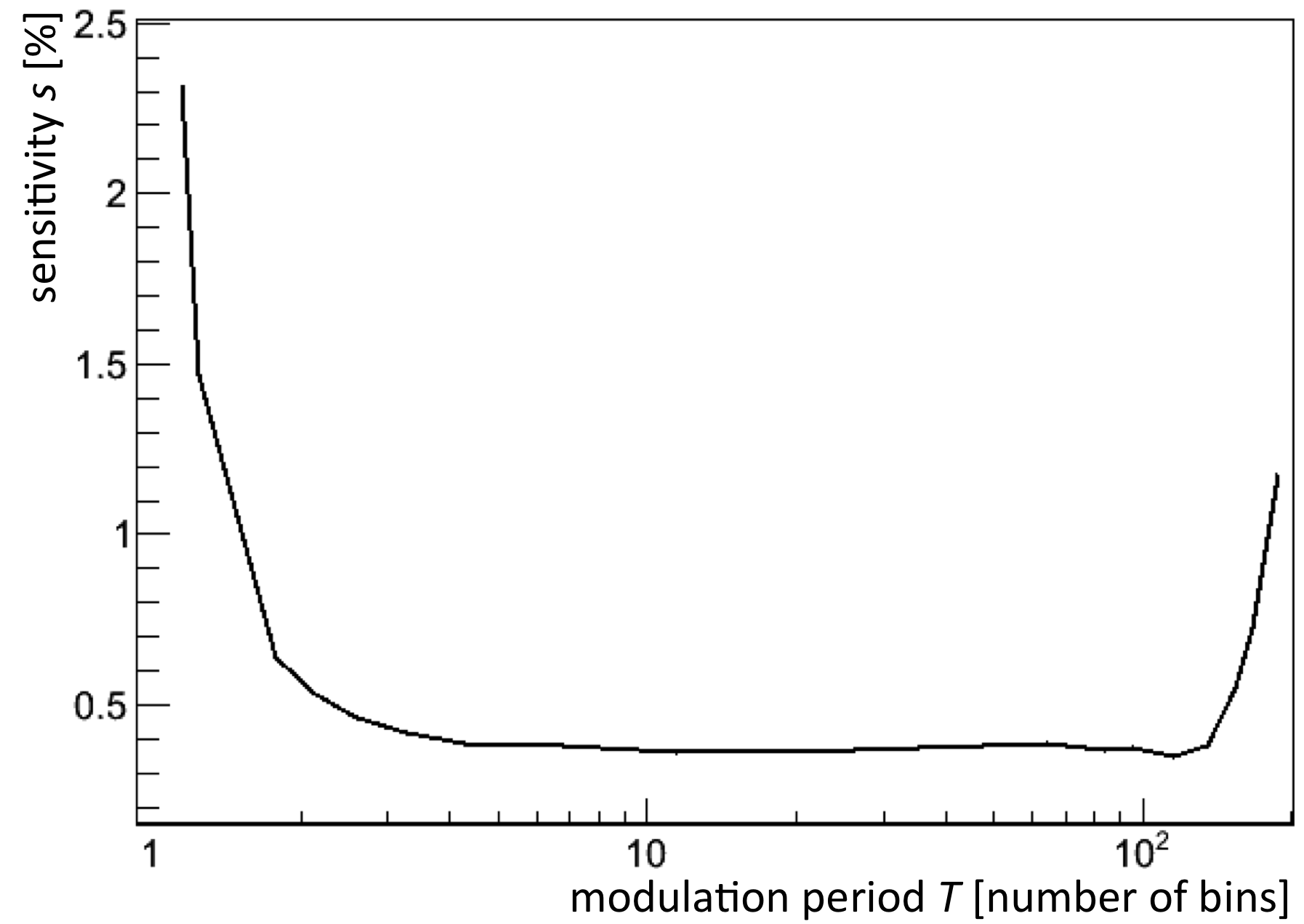}
\caption{Sensitivity dependence on modulation period $T$ versus number of bins $n$. The plot is generated assuming $n=100$ and $\Delta=1\,\%$. The minimum period $T_\mathrm{min}$ that still provides an optimum sensitivity corresponds to the width of five bins.}
\label{FigOptPer}
\end{figure}

The LS method loses in sensitivity $s$ in the case that the modulation period $T$ gets short compared to the bin width $w$. Fig.\,\ref{FigOptPer} shows $s$ for a 100 bin MC data set with a bin content uncertainty $\Delta=1\,\%$. As mentioned already in Sect.\,\ref{SecMedMod}, the sensitivity is almost constant over a wide range of periods, $5w<T<150w$. The minimum period detectable at optimum $s$ is therefore
\begin{equation}\label{EqMinPer}
T_\mathrm{min}\approx5\,w.
\end{equation}

\subsubsection*{Statistical Uncertainties for Short Time Binnings}

As shown in Sect.\,\ref{SecMedMod}, the {$^7$Be}-$\nu$ rate in LENA is best determined by a spectral fit for bin widths $w>1$\,day. However, this method will fail for $w\ll1$\,day as at some point the fit will return less precise results than a simple counting of the events inside the bin. If this counting approach is chosen, the background contributions inside the bin must be statistically subtracted from the total bin content $N_\mathrm{tot}$ in order to obtain the signal rate. As discussed in Sect.\,\ref{SecMedMod}, the primary background contributions are {$^{85}$Kr} and {$^{210}$Bi}; {$^{210}$Po} can be removed by pulse-shape discrimination (or decay). The number of {$^7$Be}-$\nu$ events in a single bin, $N_\nu(w)$, is therefore given by:
\begin{eqnarray}
N_\nu(w) = N_\mathrm{tot}(w)-N_\mathrm{bg}(w), 
\end{eqnarray}    
where $N_\mathrm{bg}=N_\mathrm{Kr}+N_\mathrm{Bi}$. The number of background events scales with the background rate $r_\mathrm{bg}$ and the bin width $w$, $N_\mathrm{bg}=r_\mathrm{bg}w$. The total number of events is $N_\mathrm{tot}=(r_\nu+r_\mathrm{bg})w$. In the following assessment, background rates are taken corresponding to the situation in Borexino \cite{bx07be7}: The signal-to-background ratio inside the detection window is roughly 1:1, so the rates in LENA are
\begin{eqnarray}\label{eqrates}
r_\nu\approx r_\mathrm{bg} \approx 10^4/\mathrm d.
\end{eqnarray}
The contribution of background events to $N_\mathrm{tot}$ varies according to the statistical uncertainty $\sqrt{N_\mathrm{bg}}$. Moreover, the statistical subtraction of $N_\mathrm{bg}$ from $N_\mathrm{tot}$ is based on $r_\mathrm{bg}$. The mean rate is determined by a spectral fit applied to a larger data sample, collected over a time span in which the contamination remains unchanged. Very conservatively assuming that the value of $r_\mathrm{bg}$ is determined on the basis of 1-day data sets, the corresponding uncertainty is of the order of $3\,\%$ (as it is the case for {$^7$Be}-$\nu$'s of Sect.\,\ref{SecMedMod}). Therefore, the total uncertainty on the neutrino rate per bin, $\Delta N_\nu(w)$, is determined by the quadratic sum of the uncertainty for the total number of events, $\Delta N_\mathrm{tot}=\sqrt{N_\mathrm{tot}}$, and for the background contribution $\Delta N_\mathrm{bg}=\sqrt{N_\mathrm{bg}+(0.03N_\mathrm{bg})^2}$:
\begin{eqnarray}
\Delta N_\nu = \sqrt{(\Delta N_\mathrm{tot})^2+(\Delta N_\mathrm{bg})^2} = \sqrt{N_\mathrm{tot}+N_\mathrm{bg}+(0.03N_\mathrm{bg})^2} = \sqrt{(r_\nu+2r_\mathrm{bg})w+(0.03r_\mathrm{bg})^2w^2}.
\end{eqnarray}
By entering the rates quoted in (\ref{eqrates}), the relative uncertainty of the {$^7$Be}-$\nu$ rate per bin can be written as:

\begin{eqnarray} \label{EqRelUnc}
\Delta_\nu=\frac{\Delta N_\nu}{N_\nu} = \frac{\Delta N_\nu}{r_\nu w}=10^{-2}\sqrt{\frac{2.9\,\mathrm d}{w}+9}.
\end{eqnarray}

\begin{figure}
\centering
\includegraphics[width=0.485\textwidth]{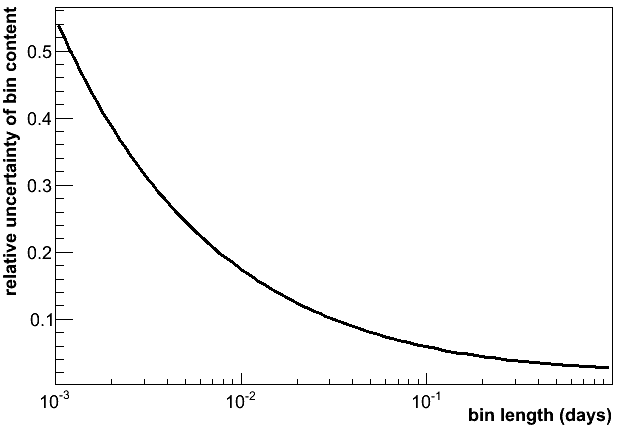}
\caption{The relative uncertainty of the bin content of {$^7$Be} events $N_\nu$ depending on the bin width $w$. The numbers are based on the statistical subtraction of the background rates caused by {$^{85}$Kr} and {$^{210}$Bi} decays.}
\label{FigUncSho}
\end{figure}

For $w=1$\,d, $\Delta_\nu=\Delta N_\nu/N_\nu\approx3\,\%$ is governed by the uncertainty in the determination of the background rate. But for bin width $w<0.3$\,d, the statistical uncertainties of the individual bins will dominate the uncertainties of the long-term background rate determinations. Fig.\,\ref{FigUncSho} shows the corresponding function of the uncertainty, $\Delta_\nu(w)$.

\subsubsection*{Single-Bin Uncertainty vs Number of Bins}

\begin{figure}
\centering
\includegraphics[width=0.485\textwidth]{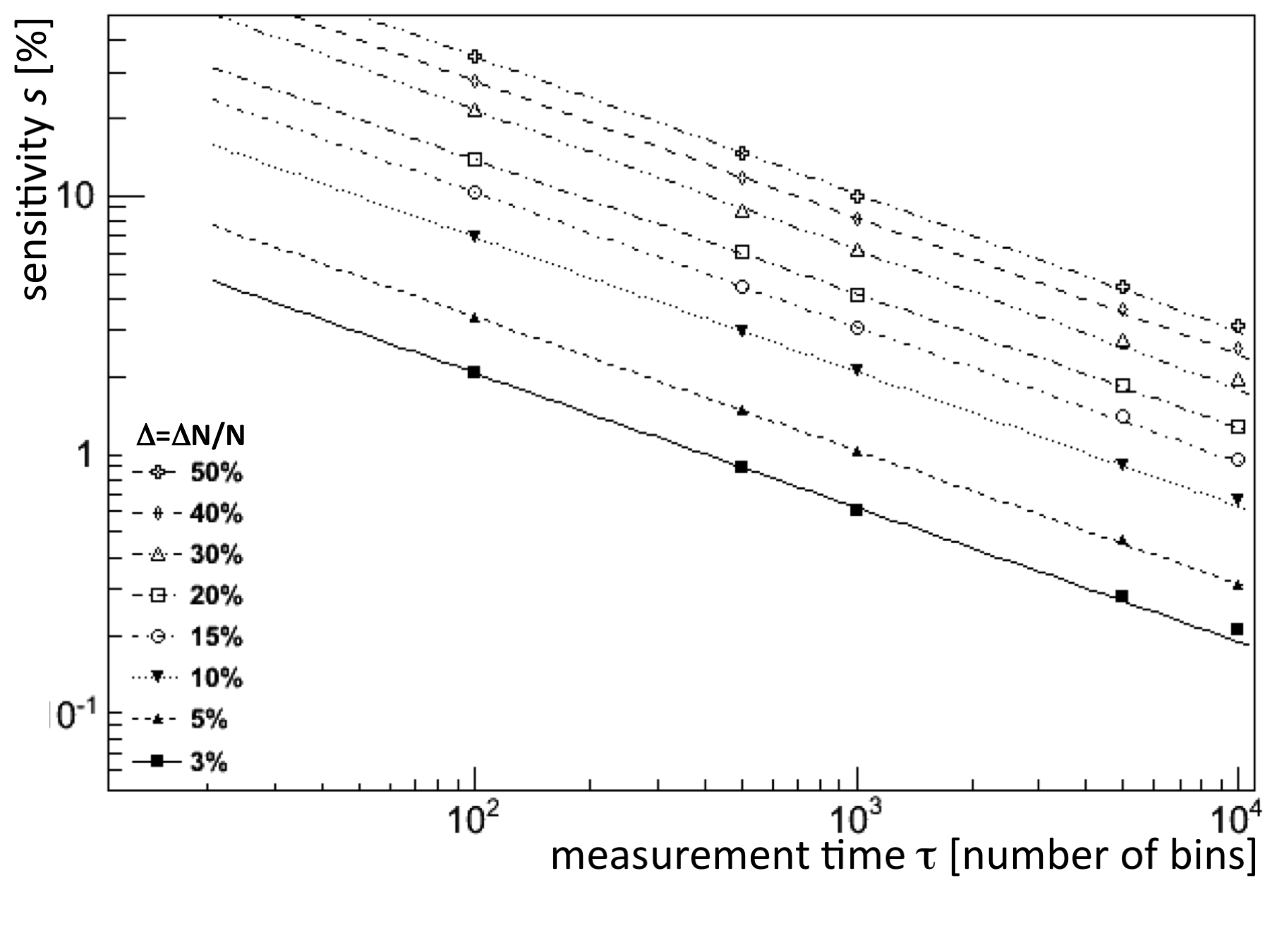}
\caption{Dependence of the sensitivity $s$ ($T_\mathrm{min}\geq5w$) on the number of bins $n$ available for LS analysis. The markers represent the results obtained by  simulation, varying both the bin-content uncertainty $\Delta$ and $n$ in the MC data sets. The lines correspond to fits to the MC results at  distinctive values of $\Delta$, following Eq.\,(\ref{EqBinSen}). Parameters are listed in Tab.\,\ref{TabTwoPar}.}
\label{FigMinBin}
\end{figure}

A fixed relation between the sensitivity $s$ and the minimum measurement time $\tau$ can be obtained by combining the bin content uncertainties $\Delta$ and the number of bins $n$ for the LS analysis. In accordance to Eq.\,\ref{EqMinPer}, only sufficiently long periods $T_\mathrm{min}\geq5\,w$ are considered. Fig.\,\ref{FigMinBin} shows the results of MC simulations describing the dependence of $s$ on $n$. As one would expect, the sensitivity of the analysis improves with $n$. In addition, the uncertainty of the bin content $\Delta$ was varied. From the MC result, it is possible to derive a function describing the dependence of $s$ on $n$ and $\Delta$:
\begin{eqnarray}\label{EqBinSen}
s(n,\Delta) = m\cdot\Delta\cdot n^{-\gamma}.
\end{eqnarray} 
The actual fit parameters vary for the individual values of $\Delta$ and are listed in Tab.\,\ref{TabTwoPar}. Approximately, the sensitivity can be described by
\begin{eqnarray}\label{EqBinSenAppr}
s(n,\Delta) \approx 7.5\cdot\Delta\cdot n^{-0.5}.
\end{eqnarray} 

\begin{table}
\begin{center}
\begin{tabular}{|c|rr|}
\hline
$\Delta$ & $m$ & $\gamma$ \\
\hline
0.03 & 7.4847(13) & 0.51880(3)   \\
0.05 & 7.3248(4) & 0.516292(13) \\
0.1  & 7.5843(8) & 0.520261(19) \\
0.15 & 7.550(2) & 0.52059(5)   \\
0.2  & 7.4675(15) & 0.51792(4)   \\
0.3  & 8.605(2) & 0.5403(5)    \\
0.4  & 7.875(7)  & 0.52803(18)  \\
0.5  & 7.938(6)  & 0.53077(14)  \\
\hline
\end{tabular}
\caption{Fit parameters for the sensitivity $s$ as function of the relative bin content uncertainty $\Delta$ and the number of measurement bins $n$, following Eq.\,(\ref{EqBinSen}).}
\label{TabTwoPar}
\end{center}
\end{table}

\subsubsection*{Sensitivity Estimate}

The relation shown in Eq.\,(\ref{EqBinSen}) describes the dependence of the sensitivity $s$ of a modulation search on the bin content uncertainty $\Delta$ and the number of bins $n$. For estimating the final sensitivity of LENA, the interesting parameters are the minimum modulation period $T_\mathrm{min}$ detectable as a function of the overall measurement time $\tau$.

\begin{figure}
\centering
\includegraphics[width=0.485\textwidth]{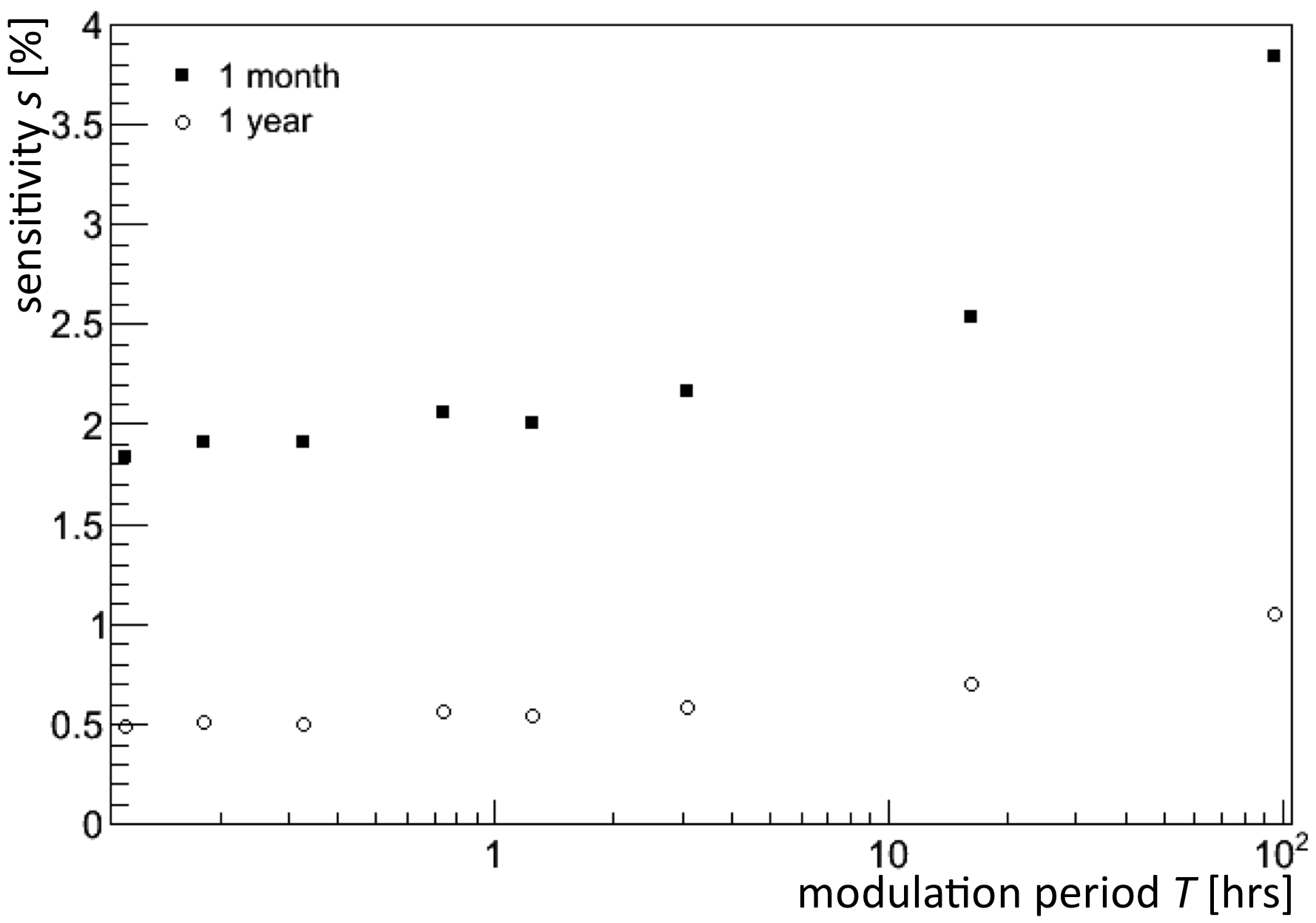}
\caption{Sensitivity $s$ in LENA for short modulation periods. The plot is generated using the dependence of sensitivity on the number of bins of Eq.\,(\ref{EqBinSen}). Each data point corresponds to an individual bin uncertainty and the correlated minimum modulation period that can be detected. For small periods, the greater number of bins is leveled by the increase in bin uncertainty and the correlated loss in sensitivity.}
\label{FigShoMod}
\end{figure}

As stated in Eq.\,(\ref{EqMinPer}), $T_\mathrm{min}$ has to include at least 5 bins. The relation between $\Delta$ and the chosen bin width $w$ is described by Eq.\,(\ref{EqRelUnc}). The measurement time $\tau$ is the product of bin width and number of bins, $\tau=w\cdot n$. The result of combining this information with Eq.\,(\ref{EqBinSen}) is depicted in Fig.\,\ref{FigShoMod}: The best sensitivity $s$ achievable is $s\approx2\,\%$ for a measurement time of 1 month, and $s\approx0.5\,\%$ for $\tau=1$\,yr. The worsening of $s$ with increasing $T$ visible in Fig.\,\ref{FigShoMod} is based on the shape of Eq.\,(\ref{EqRelUnc}): For large $T$, $\Delta N$ is dominated by the constant term originating from the long-term background fit. However, it can be avoided by choosing smaller bin widths $w\ll T$ even in the case of larger modulation periods, as Eq.\,(\ref{EqMinPer}) only provides a lower limit. Therefore, the minimum sensitivities quoted above are constant as long as $T\leq\tau$ (compare Fig.\,\ref{FigOptPer}).

The sensitivity levels obtained by this method would surpass the values presented in Sect.\,\ref{SecMedMod} provided sufficient measurement time can be reached. If one considers a data set of $n=3.65\cdot10^5$ bins of $w=10^{-2}$\,d in width (corresponding to 10 years of measurement), the corresponding bin uncertainty is $\Delta_\nu \approx 17.3\,\%$ from Eq.\,(\ref{EqRelUnc}). In this case, Eq.\,(\ref{EqBinSenAppr}) returns a sensitivity $s\approx0.2\,\%$, close to the maximum sensitivity $s=0.1\,\%$ obtained if systematical errors are not considered (corresponding to $\Delta=1\,\%$ in Tab.\,\ref{TabYrsUnc}). However, the presented analysis relies on the uniformity of both detector response and background levels over the whole measurement period. The experience of Borexino shows that this will be hard to achieve on the scale of several years. Shorter stability periods can be much easier accommodated in an analysis following Sect.\,\ref{SecMedMod}.

\section{Applications and Discussion}
\label{SecDiscus}

Based on the sensitivity estimates presented in Sect.\,\ref{SecAnaRes}, one can define three modulation period ranges that will be accessible in LENA:
\begin{itemize}
\item $10\,\mathrm{min}<T<1\,\mathrm d$: For very short periods, the LS analysis relies on simple counting of events inside the neutrino detection window. The sensitivity mainly depends on the measurement time $\tau$: For $\tau=1$\,yr, $s\approx0.5$\,\% in amplitude can be reached.
\item $1\,\mathrm d < T < 10\,\mathrm{yrs}$: For medium-range periods, $s\approx0.3$\,\% can be achieved for a measurement time of 10\,yrs based on fits to the electron recoil spectrum in day-long bins.
\item $T > 10\,\mathrm{yrs}$: For long periods, the sensitivity decreases with $T$, and is also subject to the strong influence of the phase shift $\varphi$ relative to the acquisition time window. Periods beyond a few hundred years are generally not accessible.
\end{itemize}
In the following, several possible sources of modulations in the solar neutrino flux and the sensitivity of LENA for their detection will be discussed.

\subsubsection*{Keplerian Motion}

Although already discovered in preceding solar neutrino experiments, an analysis would return a confirmation of the annual flux modulation caused by the ellipticity of the terrestrial orbit at high significance (compare Tab.\,\ref{TabYrsUnc}). For short measurement times, $\tau\approx1$\,yr, it could be argued that the presence of the large signal might cause a decrease in sensitivity for neighboring frequencies in the LS periodigram. However, MC simulations investigating this effect show that an artificial suppression of this signal in a data set by a bin-wise division with $A(t)\approx0.04\sin(2\pi (t/1\,\mathrm{yr}))$ (see Eq.\,(\ref{EqBasMod})) compensates the distortions in the baseline of the periodigram.

\subsubsection*{Day-Night Effect}
\label{SubsDayNig}

The observation of a day-night asymmetry in the solar neutrino flux, induced by the transition of solar neutrinos through the Earth's matter, has been considered as a ''smoking gun'' evidence for the MSW effect \cite{car86}. Formerly, the LOW solution for neutrino oscillation parameters predicted rather large amplitudes of up to 40\,\% for the observable {$^7$Be} flux \cite{bah97,deg01}. Due to the consolidation of the MSW-LMA model of neutrino oscillations, such a large effect seems now very unlikely. Alternatively, a similar day/night asymmetry of the survival probability is predicted in the case of mass varying neutrinos: the expected amplitude is $\sim$20\,\%, while the phase is opposed to regular matter effects \cite{deh08}. Up to now, no indication of modulations at this scale has been observed Borexino data. Nevertheless, it might be worthwhile to search for such a modulation in LENA {$^7$Be} data. Due to the long-term stability of the effect, sensitivities of $10^{-3}$ could be reached for a diurnal modulation.

\subsubsection*{Correlations to the Solar Cycle}
\label{SubsGalExp}

A variety of authors has analyzed the data collected over long measurement periods in the radiochemical experiments, searching for low-frequency modulations with periods of the order of tens of years (e.g.\,\cite{stu08}). In the data reported by all gallium experiments (Gallex/GNO and SAGE), a decrease in the neutrino flux on the order of 20\,\% over about 10 years of combined measurement time is visible \cite{gno05,sage09}. However, the statistical uncertainties on the individual measurement bins are too low to provide a significant result; data is as well consistent with a constant neutrino rate.

In terms of the analysis presented above, the inconclusive change in flux could be approximated by a modulation of $T\geq20$\,yrs  and $A=10\,\%$. Confirmation or rejection by LENA is therefore well within reach, even if the actual measurement period were only several years. However, an evaluation of the periodicity of the signal would probably require a measurement time of the order of 10 years.

\subsubsection*{Helioseismic Waves in the Neutrinosphere}
\label{SubsHelMod}

It has been argued that helioseismic waves might give rise to modulations in the solar neutrino production rate \cite{sno09}. These waves are present throughout the Sun and stretch over a large range of amplitudes and frequencies \cite{app09}. Pressure-driven waves (p-modes) propagate close to the solar surface and have been extensively studied by optical helioseismology. However, buoyancy or gravity-driven waves (g-modes) are confined to the inner regions of the Sun due to their dampening in the convective zone. Up to now, there is no unambiguous evidence for g-modes by standard helioseismology, although hints of an oscillation at a frequency of 220.7\,\textmu Hz have been found in SOHO data \cite{jim09}. 

The expected frequency range of solar g-modes can be derived from solar model calculations: Typical values are 200\,\textmu Hz or lower, corresponding to periods longer than 1.4\,h. The current best limit for a corresponding modulation in the observed neutrino flux has been derived from SNO data and excludes amplitudes of more than 10\,\% \cite{sno09}. Based on the present study, LENA will be able to surpass the accuracy of this measurement by at least a factor of 20 in the relevant frequency region.

There is no specific mechanism that relates g-mode amplitudes to a modulation of the neutrino flux. However, the primary possibilities are a change of neutrino production rate induced by a localized change in temperature, or a variation of the matter potential influencing the MSW survival probability (see below). 

In the first case, the sensitivity of the search would be enhanced by the steep dependence of the fusion rate on the temperature of the medium \cite{bah05}. Standard solar model calculations indicate that the {$^7$Be} rate scales with the 11th power of solar core temperature. This implies that relative temperature changes on the level of $s/11\approx5\times10^{-4}$  might be detectable in LENA. The variation will be even more pronounced in the {$^8$B} fusion rate, as the temperature dependence $\phi_\mathrm{^8B}\propto{T_\mathrm{c}}^{25}$ is even steeper. The considerably lower sensitivity (compared to that of {$^7$Be}-$\nu$) to {$^8$B}-$\nu$ flux modulations of $s\approx 1\,\%$ in LENA, corresponding to a detection rate of $\sim$100\,cpd, would be roughly compensated by the enhanced temperature sensitivity. 

Helioseismic waves are usually quantified by means of spherical harmonics. Solar model calculations indicate the largest oscillation amplitude for g-modes of low quantum numbers \cite{app09}. This is fortunate, as the full effect of temperature variation will only be observed if all of the {$^7$Be} (or {$^8$B}) fusion region undergoes the oscillation in unison.
\subsubsection*{Solar Matter Effects and Spin Conversion}

The influence of the solar matter on the survival probability of {$^8$B} neutrinos is a basic ingredient of the MSW-LMA oscillation model. However, the predictions differ concerning the question whether changes in solar density, as induced by helioseismic activity, are sufficiently pronounced to influence the $\nu_e$ survival probabilities \cite{bur02}.

While LENA's primary sensitivity for modulation search is in the {$^7$Be} signal, it seems very unlikely that matter effects will have a significant influence on the survival probability at sub-MeV energies. However, the {$^8$B} neutrino rate will be sensitive to such effects. For a rate of 100\,cpd, a minimum amplitude of 1\,\% could be identified. This surpasses the current best limit held by SNO by an order of magnitude \cite{sno09}.

Another possible cause for modulations in the solar $\nu_e$ flux is the conversion of neutrino flavors, or even neutrino-antineutrino conversion, based on the interaction of a non-zero magnetic moment of neutrinos with the solar magnetic field \cite{sch81}. The flavor conversion was originally an alternative oscillation mechanism to mass-induced oscillations and is disfavored as the dominant effect by experimental results. However, an analysis monitoring the $\nu_e$ survival probability as a function of the solar magnetic field strength is likely to return similar sensitivities as the periodic modulation searches. It will also be sensitive to the $\nu\rightarrow\bar\nu$ conversion.

For the latter case, both KamLAND and Borexino have performed direct searches for the appearance of $\bar\nu_e$ events in the energy region above 8\,MeV, in which the overwhelming background of reactor $\bar\nu_e$ declines \cite{kam05an,bx10}. It is obvious that LENA could put much more stringent limits due to its large target mass. However, also in this energy region there is an irreducible background due to the $\bar\nu_e$ component of the diffuse Supernova neutrino background, which poses a natural limit \cite{wur07dsn}. 

\section{Conclusions}
The next generation of large-volume, low-energy neutrino detectors will enhance the experimental sensitivity towards modulations in the solar neutrino flux \cite{tur10}. A liquid-scintillator detector seems especially attractive due to the possibility to observe the {$^7$Be} neutrino line at 866\,keV: As the expected event rate is of the order of $10^4$ counts per day, the high statistics will allow to search for modulations on a sub-percent level, by far surpassing the sensitivity of currently running experiments. The sensitivity of the search stretches from very short time scales of the order of tens of minutes to tens or even hundreds of years. This will allow to probe the frequency regions of the helioseismic observation for g-mode oscillations in the solar center \cite{app09}, the day-night effect \cite{bah97,deg01}, and to search for variations of the fusion rate with the solar cycle. 

\section*{Acknowledgements}
This work was supported by the Maier-Leibnitz-Laboratorium (Garching), the Deutsche Forschungsgemeinschaft DFG (Transregio 27: Neutrinos and Beyond), and the Munich Cluster of Excellence ''Origin and Structure of the Universe''. We would like to thank the Borexino collaboration for the common work and many fruitful discussions.

\bibliographystyle{h-physrev}
\bibliography{solar_arxiv}

\end{document}